\documentclass[10pt,letterpaper]{article}
\usepackage[top=0.85in,left=2.75in,footskip=0.75in,marginparwidth=2in]{geometry}

\usepackage[utf8]{inputenc}

\usepackage{xcolor}

\usepackage{nameref,hyperref}

\usepackage[right]{lineno}

\usepackage{microtype}
\DisableLigatures[f]{encoding = *, family = * }

\raggedright
\usepackage{changepage}

\usepackage[aboveskip=1pt,labelfont=bf,labelsep=period,singlelinecheck=off]{caption}

\usepackage[round]{natbib}

\usepackage{amsfonts,amsmath,amssymb}

\makeatletter
\renewcommand{\@biblabel}[1]{\quad#1.}
\makeatother

\usepackage{lastpage,fancyhdr,graphicx}
\usepackage{epstopdf}
\fancyheadoffset[L]{2.25in}
\fancyfootoffset[L]{2.25in}

\usepackage{color}

\definecolor{Gray}{gray}{.25}

\usepackage{graphicx}

\usepackage{sidecap}

\usepackage{wrapfig}
\usepackage[pscoord]{eso-pic}
\usepackage[fulladjust]{marginnote}
\reversemarginpar

\begin{document}
\vspace*{0.35in}

\begin{flushleft}
\begin{adjustwidth}{-1.5in}{0in} 
\small{\textcolor{gray}{This version of the paper is submitted to the Quarterly Journal of the Royal Meteorological Society.}}
\end{adjustwidth}
{\Large
\textbf\newline{Excitation of mixed Rossby-gravity waves by wave - mean flow interactions on the sphere}
}
\newline
\\
Sándor István Mahó\textsuperscript{1,*},
Sergiy Vasylkevych\textsuperscript{1},
Nedjeljka \v{Z}agar\textsuperscript{1},
\\
\bigskip
\bf{1} Meteorological Institute, Center for Earth System 
Research and Sustainability, Universit\"at Hamburg, Hamburg, Germany
\\
\bigskip
* sandor.maho@uni-hamburg.de
\bigskip

\end{flushleft}

\section*{Abstract}
The equatorial mixed Rossby-gravity wave (MRGW) is an important contributor to tropical variability. Its excitation mechanism capable of explaining the observed MRGW variance peak at synoptic scales remains elusive. This study  investigates wave-mean flow interactions as a generation process for the MRGWs using the barotropic version of the global Transient Inertia-Gravity And Rossby wave dynamics model (TIGAR), which employs Hough harmonics as the basis of spectral expansion, thereby representing MRGWs as prognostic variables. High accuracy numerical simulations manifest that interactions between waves emanating from a tropical heat source and zonal mean jets in the subtropics generate MRGWs with the variance spectra resembling the one observed in the tropical troposphere. Quantification of spectral tendencies associated with the MRGW energy growth underscores the significance of wave-mean flow interactions in comparison to excitation mechanisms driven by external forcing and wave-wave interactions. The MRGW growth and amplitude depend on the asymmetry in the zonal mean flow that may explain not only seasonal variability but also differences between the troposphere and the middle atmosphere.

\section{Introduction}
Since \citet{yanai1966} first discovered the mixed Rossby-gravity waves (MRGWs or the MRG modes) as periodic fluctuations in the meridional wind in the equatorial stratosphere, their properties and the associated excitation mechanism have been a subject of numerous studies. Theoretically, the MRGW, also known as Yanai wave, is a westward-propagating solution of the linearized shallow water equations on the equatorial $\beta$-plane as derived by \citet{matsuno1966}.

Initially, the MRGWs were analysed as free modes, however, satellite irradiance data soon provided evidence that tropospheric MRGWs are coupled to convection \citep{chang1970,wallace1972,hendon1991,takayabu1994,bergman1994, wheeler_kiladis1999}. The MRGWs were argued to provide energy for tropical depression type waves over the equatorial Pacific \citep{takayabu1993} that can trigger the formation of tropical storms \citep{dickinson2002_TD_wave}. In addition, their vertical propagation was shown to contribute to the forcing of the easterly phase of the quasi-biennial oscillation (QBO) \citep{lindzen_holton1968,alexander2008global,kawatani2010}. Furthermore, MRGWs were shown to contribute to the MJO initiation \citet{takasuka2019}. In a recent study, tropospheric MRGWs were associated with the Hadley cell dynamics \citep{hoskins2021}. Observational studies agree about the dominant zonal wavenumbers of 4-5 of the MRGWs and periods of about 5 days \citep[e.g.][]{yanai1969large,yanai1970further,zangvil1980}.

Early studies of the MRGW excitation mechanism applied the adiabatic barotropic or baroclinic models and found that MRGWs can be generated by long-period, large-scale tropospheric heating perturbations asymmetric about the equator \citep{holton1972waves_heatsrc}. This excitation process is easily demonstrated but can not explain the observed scale selection. Other excitation mechanisms that have been studied in depth include the lateral boundary forcing \citep{mak1969}, the wave-CISK (conditional instability of the second kind) mechanism \citep{hayashi1970} and the wave interactions \citep{raupp2005}.

The theory proposed by \citet{mak1969} and refined by \citet{wilson_mak1984} states that the scale selection of MRGWs is driven by a linear resonance mechanism that is directly controlled by the properties of the forcing wave exerted at lateral boundaries.  \citet{zhang1993} later showed that the amplitude of laterally forced MRGWs changes with the mean flow due to the Doppler-shift effect on the forcing frequency. 
\citet{hayashi1970} argued that instability of the wave-CISK mechanism, which describes interaction between cumulus convection and equatorial wave dynamics, is unstable and can produce MRGWs. However, this mechanism was shown not capable of explaining the scale selection of the MRGWs \citep{itoh1988}. As an alternative, \citet{itoh1988} suggested a non-linear wave-CISK theory in which the MRGWs are amplified by asymmetric lateral boundary forcing.
Finally, a recent study by \citet{raupp2005} suggested wave interactions as a generation mechanism for MRGWs. The authors numerically solved the non-linear shallow-water equations on the equatorial $\beta-$plane and argued that MRGWs were excited by a non-linear mechanism in which an asymmetric heat source produces waves that destabilise the basic state, which supplies energy for the development of MRGWs. 

The present study contributes to the understanding of the MRGWs by showing that they can be excited by non-linear interactions of wave perturbations
and asymmetric zonal mean flows in the tropics or subtropics. This is demonstrated by a hierarchy of high-accuracy numerical simulations using idealized and observed latitudinal profiles of the zonal mean zonal wind.

The new excitation mechanism of MRGWs is presented using the recently developed barotropic version of the TIGAR model (Transient Inertia-Gravity And Rossby wave dynamics). TIGAR \citep{Vasylkev2021} solves the rotating shallow water (RSW) equations on the sphere and applies Hough harmonics as spectral expansion basis. In this framework, the MRGW along with other eigenmodes of the linearized system (i.e. Kelvin waves, inertia-gravity modes and Rossby modes on the sphere) is a prognostic variable represented by a spectral Hough expansion coefficient for each zonal wavenumber. In other words, the MRG wave is defined by the projection of height and winds onto the spatial structure of the corresponding mode on the sphere and processes contributing to its tendencies are quantifiable. MRGW structures have geopotential height and zonal wind field asymmetric with respect to the equator and symmetric meridional wind field with equatorial maximum\footnotemark[1] \footnotetext[1]{The horizontal structure of the zonal wavenumber 1 MRGW for the mean depths from 10 km to about 27 meters is shown at  \url{https://modes.cen.uni-hamburg.de/Hough##part2_2}}. Our numerical simulations demonstrate that the subtropical jet has a prominent role in wave$-$mean flow interactions that leads to the excitation of the MRG mode. 

The remainder of this paper is structured as follows. In Section \ref{sec2} we describe the model and its setup for the hierarchy of numerical experiments and demonstrate the model capabilities. The results are presented in the two sections focusing on the non-linear excitation in simulations with idealized (Section \ref{sec4}) and realistic (Section \ref{sec5}) profiles of the background zonal wind. The discussion and conclusions are presented in Section \ref{sec6}.

\section{Numerical model formulation and setup}\label{sec2}

We use the 2D version of the TIGAR model, which was introduced in \citet{Vasylkev2021}. Below we recall its formulation, which is followed by a description of the linearized version of TIGAR, which is developed for the purpose of this study. The list of numerical experiments concludes the section.

\subsection{The non-linear TIGAR model}

The model equations in spherical coordinates $(\lambda,\varphi) \in [0,2\pi) \times (-\pi/2,\pi/2)$ read as follows:

\begin{subequations} \label{eq:RSW}
\begin{gather}
    \frac{\mathrm{d} u} {\mathrm{d} t} - v \left(2\Omega \sin{\varphi} +\frac{u}{a}\tan{\varphi} \right)+ \frac{g}{a\cos{\varphi}} \frac{\partial h}{\partial \lambda} = -\frac{1}{\tau_R}u-F_u \,, \\
    \frac{\mathrm{d} v} {\mathrm{d} t} + u \left(2\Omega \sin{\varphi}+\frac{u}{a}\tan{\varphi}\right) + \frac{g}{a} \frac{\partial h}{\partial \varphi} = -\frac{1}{\tau_R}v-F_v \,, \\
     \frac{\mathrm{d} h} {\mathrm{d} t} + h \nabla \cdot \textbf{V}=Q-\frac{1}{\tau_N}\left(h - D \right)-F_h \,,
\end{gather}
\end{subequations}
where $\frac{\mathrm{d}}{\mathrm{d} t}$ is the material derivative,  $\textbf{V}=[u,v]$ denotes the horizontal velocity vector composed of the zonal ($u$) and the meridional component ($v$), $h$ is the total fluid depth, $D$ is the mean depth, $g$ and $\Omega$ are the gravity constant and the rotation rate of the Earth respectively, $a$ is the radius of the Earth. The spectral viscosity denoted as $\textbf{F}=[F_u,F_v,F_h]$ acts similar to the one proposed by \citet{gelb2001} by dampening the smallest resolved scales in Equation \ref{eq:RSW} in spectral space. The heat source $Q$ is modelled as a Gaussian, as explained in Section \ref{setup} Two other dissipative terms, the so-called  Rayleigh friction in the momentum equations and the Newtonian cooling in the continuity equation, are defined by their characteristic time scales  $\tau_R$ and $\tau_N$, respectively. 

TIGAR solves Equation \ref{eq:RSW} by applying the Hough harmonics as basis functions for the spectral expansion. The Hough harmonics are the eigensolutions of the linearized rotating shallow-water equations \citep{longuet1968hough}, and they involve three types of solutions: the eastward propagating inertia-gravity modes (EIG), the westward propagating inertia-gravity modes (WIG), and the low-frequency Rossby modes. The complete set of eigenmodes involves two special solutions: the westward propagating equatorial mixed Rossby-gravity and the eastward propagating equatorial Kelvin mode, which fill the frequency gap between the slow Rossby and fast inertia-gravity (IG) modes. 
The solution of Equation \ref{eq:RSW} in terms of the Hough harmonics was proposed by \citet{kasahara1977} and TIGAR follows the steps therein. First, the non-dimensional variables $(\Tilde{u},\Tilde{v},\Tilde{h})$ are expanded in terms of the Hough harmonics as
\begin{equation}
    [\Tilde{u},\Tilde{v},\Tilde{h}]^T (\lambda,\phi,\Tilde{t}) = \sum_{k,n,l} \ \chi_{n,l}^k (\Tilde{t})  \ \textbf{H}_{n,l}^k(\lambda,\phi) \, ,
    \label{eq:sp_exp}
\end{equation}
where $\chi_{n,l}^k$ is the Hough expansion coefficient and $\textbf{H}_{n,l}^k$ is the Hough harmonic with the zonal wavenumber $k$, meridional mode index $n$ and the type of mode $l$. The latter takes values 1, 2 and 3 for the Rossby, EIG and WIG modes, respectively. Both $k$ and $n$ start with 0. The set of Hough harmonics is uniquely (up to a multiplicative constant) defined, except for  $k=0$ Rossby modes, which are the zonal steady states of linearized RSW equations. For the latter we use so-called K-modes derived by \citet{kasahara1978}. The MRG and the Kelvin waves correspond to $n=0$ index for the Rossby and EIG modes, respectively. The relation between the dimensional and non-dimensional variables is the following: $\Tilde{u} = u/\sqrt{gD}$, $\Tilde{v} = v/\sqrt{gD}$ and $\Tilde{h} = h/D - 1$. The non-dimensional time $\Tilde{t}$ is obtained by multiplying  $t$ by $2\Omega$.

The substitution of the expansion (Equation \ref{eq:sp_exp}) into Equation \ref{eq:RSW} leads to the spectral ODE 
\begin{equation}
    \frac{d}{d\Tilde{t}} \ \chi_{n,l}^k (\Tilde{t})+\left( d_{n,l}^k+i{\Tilde{\omega}}_{n,l}^k\right) \  \chi_{n,l}^k (\Tilde{t}) = f_{n,l}^k(\Tilde{t}) \, 
    \label{eq:sp_ODE}
\end{equation}
for the  model prognostic variables $\chi_{n,l}^k$. In Equation \ref{eq:sp_ODE} $\Tilde{\omega}$ denotes the non-dimensional frequency of the mode $(k,n,l)$. Term $d_{n,l}^k$ stands for the spectral viscosity, which is expressed by Equations \ref{eq:sp_visc}-\ref{eq:sp_visc_dist}:
\begin{equation}\label{eq:sp_visc}
d_{n,l}^k = 
\left\{
    \begin{array}{lr}
        0, & \text{if } N \leq N_c\\
        \epsilon q^2 N^2 (N + 1)^2, & \text{if } N > N_c
    \end{array}
\right\}
\end{equation}
where
\begin{equation}\label{eq:tot_wn}
    N = k + n,
\end{equation}
\begin{equation}\label{eq:sp_visc2}
    q = \exp\left(\frac{-(N-M)^2}{2(N-N_c)^2}\right),
\end{equation}
and 
\begin{equation}\label{eq:sp_visc_dist}
    \epsilon = \frac{1}{\tau_{SV} M^3} \ \ .
\end{equation}
Note that spectral viscosity only acts beyond a pre-defined cut-off total wavenumber $N_c$. For example, for T170 resolution, $N_c$ is set to the value of 94. The strength of $d_{n,l}^k$ is controlled by the dissipation time $\tau_{SV}$, which is set to 884 hours. In addition, $M$ in Equations \ref{eq:sp_visc2} and \ref{eq:sp_visc_dist} stands for the truncation limit.
We employ spectral viscosity to ensure realistic energy spectra near the truncation limit in the long simulations, while minimally impacting the dynamics otherwise. 

The right-hand side term $f_{n,l}^k$ in Equation \ref{eq:sp_ODE} contains the contribution of non-linear interactions and forcing as 
\begin{equation}
    f_{n,l}^k = \frac{1}{2\pi} \int_{0}^{2\pi} \int_{-\pi/2}^{\pi/2} \left(\textbf{N} + \mathcal{F} \right) \left( \sum_{k',\,n',\,l'} \ \chi_{n',\,l'}^{k'}  \ \textbf{H}_{n',\, l'}^{k'}  \right) \cdot \left( \textbf{H}_{n,l}^{k} \right)^* (\lambda,\varphi) \  d\varphi \ d\lambda \, ,
    \label{eq:nlin_integral}
\end{equation}
where 
\begin{equation}
    \textbf{N}= - \gamma 
    \begin{bmatrix}
     \Tilde{\textbf{V}}\cdot\Tilde{\nabla}\Tilde{u}-\Tilde{u}\Tilde{v} \ \tan(\varphi) \\ \Tilde{\textbf{V}}\cdot\Tilde{\nabla}\Tilde{v}+\Tilde{u}^2 \ \tan(\varphi) \\ \Tilde{\textbf{V}}\cdot\Tilde{\nabla}\Tilde{h}+\Tilde{h}\Tilde{\nabla}\cdot\Tilde{\textbf{V}}
    \end{bmatrix} \quad \text{and} \quad
    \mathcal{F}=
    \begin{bmatrix}
    -\frac{1}{\tau_R}\Tilde{u} \\
    -\frac{1}{\tau_R}\Tilde{v} \\
    Q-\frac{1}{\tau_N}\Tilde{h} 
    \end{bmatrix}  \, .
    \label{eq:RSW_nlin}
\end{equation}
In the shallow water setting, both the properties of linear waves and the scaling ratio between linear terms and non-linearity are controlled by a single non-dimensional quantity $1/\gamma^2  = (2a\Omega)^2/gD$  \citep{longuet1968hough,boyd_zhou_2008}, which is known as Lamb's parameter. 
The forcing and non-linear terms, $\mathcal{F}$ and $\mathbf{N}$ respectively, appearing in $f_{n,l}^k$ are computed pseudo-spectrally, i.e. they are evaluated in physical space, then transformed to spectral space by applying FFT and Hough expansion in the zonal and meridional directions, respectively. To this end Hough functions and their derivatives are precomputed with high precision from Legendre polynomials; thereby the spectral truncation is the sole source of numerical errors in the evaluation of spectral tendencies. 

The contribution of interactions between different modes to the spectral tendencies are computed directly in the model by replacing full fields in $\mathbf{N}$ by the physical space reconstruction of the modes of interest. In summary, using TIGAR for the present study has a double benefit of combining high precision computations with the ability to separate the dynamical contributions of forcing, linear dynamics, and non-linear interactions between separate modes during the model integration. Note that $\mathbf{N}$ also includes all interactions of the mean state, which correspond to $k^\prime=0$; thus, in the following, term "non-linear" will refer to both wave-wave and wave-mean flow interactions.

\subsection{Linear and linearized TIGAR}

TIGAR can be used in a linear regime that simply omits the evaluation of $\mathbf{N}$ in Equation \ref{eq:nlin_integral}. We will refer to this setup as "linear model". The linear prognostic equations are

\begin{subequations} \label{eq:RSW_lin}
\begin{gather}
    \frac{\partial u} {\partial t} - 2\Omega \sin{\varphi} v + \frac{g}{a\cos{\varphi}} \frac{\partial h}{\partial \lambda} = -\frac{1}{\tau_R}u-F_u \,, \\
    \frac{\partial v} {\partial t} + 2\Omega \sin{\varphi} u + \frac{g}{a} \frac{\partial h}{\partial \varphi} = -\frac{1}{\tau_R}v-F_v \,, \\
    \frac{\partial h} {\partial t} + \nabla \cdot \textbf{V}=Q-\frac{1}{\tau_N}\left(h - D \right)-F_h \, .
\end{gather}
\end{subequations}

"Linearized TIGAR" refers to the setup, which solves the RSW equations linearized about a prescribed geostrophically balanced mean zonal flow. Separating the model prognostic variables into the zonal mean part (denoted by an overbar) and a perturbation (denoted by '),
\begin{equation}
    u = \overline{u}(\varphi) + u'(\lambda,\varphi,t) \nonumber \,,  \quad  v = v'(\lambda,\varphi,t) \,, \quad h = \overline{h}(\varphi) + h'(\lambda,\varphi,t) \,, 
\end{equation}
the linearized model equations read
\begin{subequations} \label{eq:RSW_linzed}
\begin{gather}
    \frac{\partial u'} {\partial t} - v' \ 2\Omega \sin{\varphi} + \frac{g}{a\cos{\varphi}} \frac{\partial h'}{\partial \lambda} = -\frac{\overline{u}}{a \cos \varphi} \frac{\partial u'}{\partial \lambda} - \frac{v'}{a} \frac{\partial \overline{u}}{\partial \varphi}+\frac{\overline{u}v'}{a} \tan \varphi-\frac{1}{\tau_R}u'-F_u  \, \\
    \frac{\partial v'} {\partial t} + u' \ 2\Omega \sin{\varphi} + \frac{g}{a} \frac{\partial h'}{\partial \varphi} = -\frac{\overline{u}}{a \cos \varphi}\frac{\partial v'}{\partial \lambda}-2\frac{\overline{u}u'}{a} \tan \varphi-\frac{1}{\tau_R}v'-F_v   \, \\
    \frac{\partial h'} {\partial t} = -\frac{\overline{u}}{a \cos \varphi}\frac{\partial h'}{\partial \lambda}-\frac{v'}{a} \frac{\partial \overline{h}}{\partial \varphi}-\frac{\overline{h}}{a \cos \varphi} \frac{\partial u'}{\partial \lambda}+\frac{\overline{h}}{a} \frac{\partial v'}{\partial \varphi}+Q-\frac{1}{\tau_N}\left(h' - D \right)-F_h \,,
\end{gather}
\end{subequations}
which leads to a spectral equation similar to Equation \ref{eq:sp_ODE}, where the new terms including the means are incorporated into $f_{n,l}^k$.  Note that unlike linear TIGAR, the linearized model includes the effects of wave-mean flow interactions, while neither linear nor linearized model support wave-wave interactions. 

\subsection{Setup of numerical experiments}\label{setup}

Numerical simulations are carried out using geostrophically balanced background zonal flows, which is prescribed either analytically or derived from reanalyses. Additional simulations use the background state of rest. 
TIGAR generates balanced background flows by projecting the zonal wind field onto $k=0$ Rossby modes and solving the geostrophic balance equation in spectral space, thereby obtaining the balanced height field. The formulation of the background state accounts for the symmetry and strength of the jet. 
To compare and quantify the importance of other factors such as linear dynamics, wave-wave and wave-mean flow interactions, location and symmetry of the wave source and the mean depth, we perform a number of sensitivity experiments, which are summarized in Table \ref{t1}. Note that throughout the article the forcing is called symmetric whenever $Q$ is centred at the equator, otherwise we refer to asymmetric forcing. 

\begin{figure}[ht] 
\includegraphics[width=0.4\textwidth]{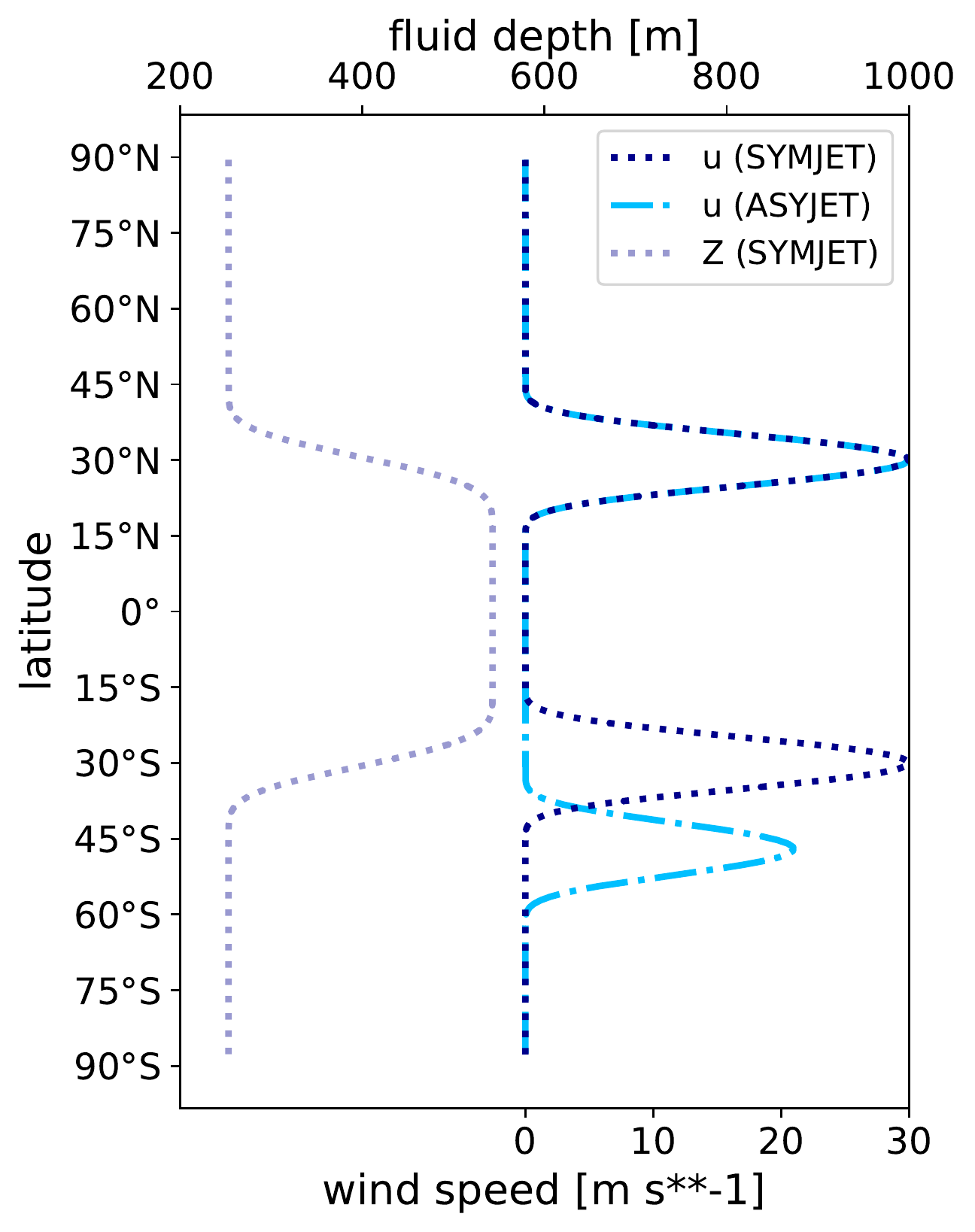}
\caption{Latitudinal profiles of the idealized zonal mean background flows (SYMJET and ASYJET). The balanced height field of SYMJET is also added.}
\label{fig:back_flows} 
\end{figure}

First, we apply idealized zonal wind profiles with a steady jet (denoted JET). The JET  simulations are further split into SYMJET and ASYJET that include two jets each located at the subtropics in both hemispheres, which are either symmetric or asymmetric with respect to the equator. These profiles are shown in Figure~\ref{fig:back_flows} along with an example of the balanced height field. The ASYJET zonal wind profile has a weaker jet in the Southern Hemisphere (SH) that is shifted further away from the equator compared to the jet north of the equator.  
Simulations with a motionless initial state are denoted NOF.

\begin{figure}[ht] 
\includegraphics[width=0.95\textwidth]{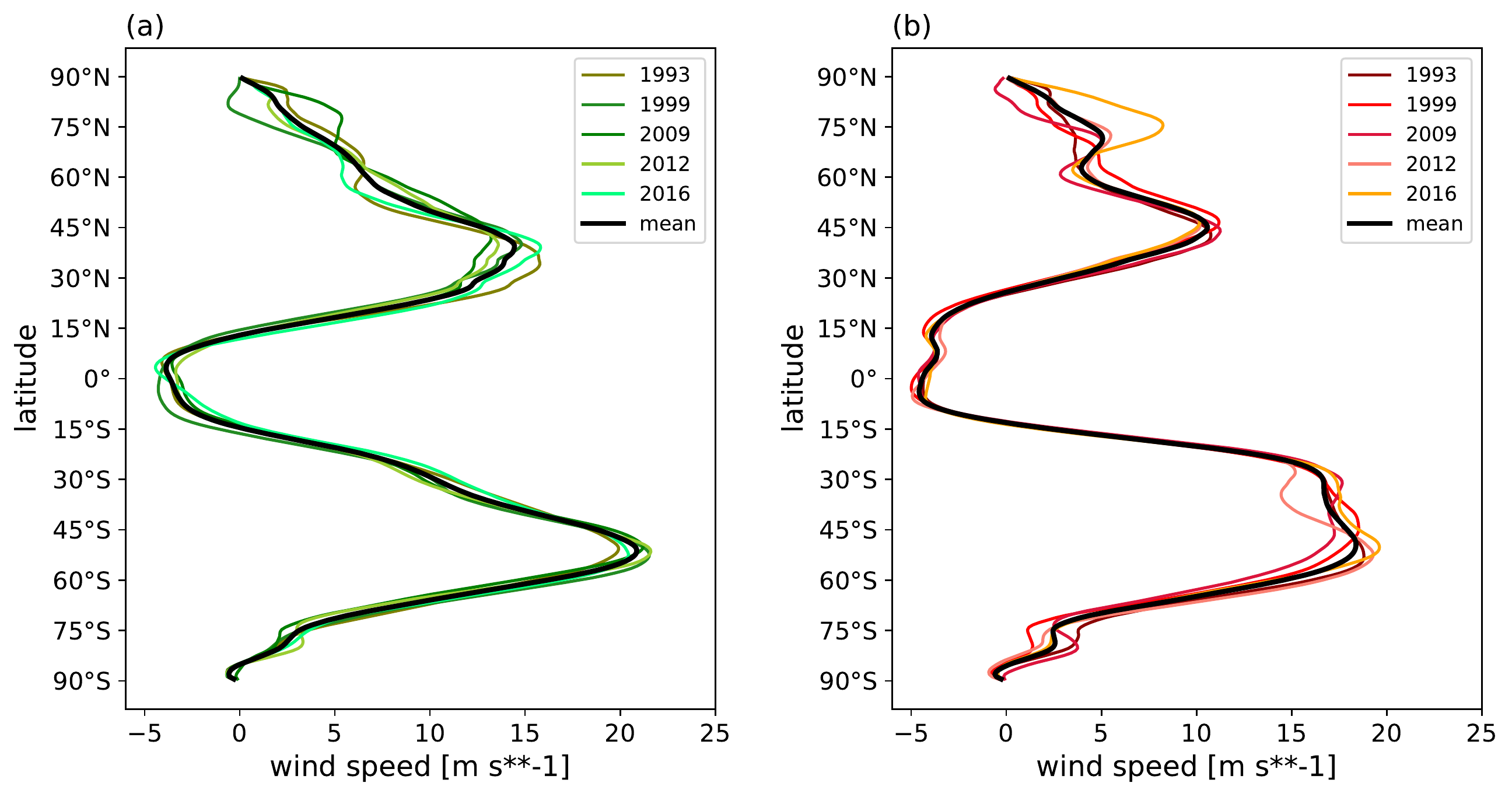}
\caption{Zonally-averaged zonal wind in ERA5 reanalysis at 500 hPa for (a) MAM and (b) JJA seasons. Five selected years are 1993, 1999, 2009, 2012 and 2016. The mean profiles are shown by solid black lines.}
\label{fig:era5_flows} 
\end{figure}

Realistic background zonal wind profiles are derived from ERA5 reanalysis data  \citep{hersbach2020era5}. We use seasonal zonally-averaged zonal wind at 500 hPa  in years 1993, 1999, 2009, 2012 and 2016 that include both strong El Nino and La Nina cases. The profiles are shown in Figure \ref{fig:era5_flows} for JJA and MAM seasons.

A visual inspection of Figure \ref{fig:era5_flows} suggests that zonally-averaged zonal wind is more asymmetric in JJA than in MAM. In order to quantify the asymmetry of the background state, we define a zonal asymmetry measure (ZAM) in the space of Hough harmonics as the ratio between energies in the asymmetric and symmetric balanced mean state (i.e., Rossby $k$=0) modes: 
\begin{equation}
     ZAM \equiv \frac{ \left| \sum_{n=0}^{M}{\chi_{2n,l=1}^{k=0}} \right|^2}{\left| \sum_{n=0}^{M} {\chi_{2n+1,l=1}^{k=0}} \right|^2} = \frac{ \left| \sum_{n=0}^{M} {\chi_{l=1}^{k=0}(asym.)} \right|^2}{\left| \sum_{n=0}^{M}{\chi_{l=1}^{k=0}(sym.)} \right|^2}  \,.
    \label{eq:ZAM}
\end{equation}
The larger the ZAM, the larger the asymmetry. The mean value of ZAM for JJA and MAM seasons is 0.321 and 0.053 respectively. The flow asymmetry is greater for the DJF season compared to SON (not shown), but we find that the differences are the largest between JJA and MAM and limit our simulations to these two seasons. 

Furthermore, we evaluate the conditions for barotropic instability of the applied  background profiles using the Rayleigh-Kuo criterion \citep{kuo1949} and find that the jet flanks of ASYJET and SYMJET profiles are unstable. In the case of the profiles derived from ERA5, the tropics and the midlatitude region appears barotropically stable, whereas the poles are moderately unstable. The difference between the stability of the zonally averaged zonal flow in MAM and JJA is negligible. 

Simulations are run with two mean depths: $D = 250$ m and $D=400$ m, which represent two characteristic values of the equivalent depth of the tropical atmosphere. The former has been commonly used for studying the tropical response to heating perturbations in adiabatic barotropic models with a lid at the tropopause \citep[e.g.][]{kasahara_silvadias1986,raupp2005}, whereas the latter is the equivalent depth of vertical mode associated with deep convection when the stratosphere is included in the computation of the vertical structure functions \citep{zagar2022_KW}. 

The forcing $Q$ is represented by a Gaussian function with the zonal and meridional scales $L_x$ and $L_y$, respectively, central location ($\lambda_0,\varphi_0$) and amplitude $A$:
\begin{equation}
    Q=A  \exp\left(-\frac{(\lambda-\lambda_0)^2}{2L_x^2}-\frac{(\varphi-\varphi_0)^2}{2L_y^2}\right) \, .
    \label{eq:heat_src}
\end{equation}
The amplitude of the forcing is $A=1.9$ m/day and 3.05 m/day for $D = 250$ m and $400$ m,  respectively, which corresponds to approximately 2.5 K/day heating rate. This is consistent with typical radiative heating rate of convective cells in the tropics \citep{mcfarlane2007}. The linear damping time scale is chosen equal for the Rayleigh friction and the Newtonian cooling, $\tau_R=\tau_N =5$ days.

Numerical simulations are performed using the triangular truncation T170 that corresponds to the F128 regular Gaussian grid. The model is integrated in time using the fourth order exponential Runge-Kutta method with time step of 5 minutes.

\begin{table}[!h]
\begin{adjustwidth}{-1.5in}{0in} 
\caption{Summary of the performed numerical experiments. The simulations are defined by the symmetry of the forcing, the shape of the background zonal wind, linearity and and the applied mean depth $D$. NL, LI and LZ refers to non-linear, linear and linearized model setup, respectively.}
\label{t1}
\begin{center}
\begin{tabular}{| c | c | c | c | c |}
\hline

\textbf{Exp label} & \textbf{Forcing symmetry} & \textbf{Background flow} & \textbf{Linearity} & \textbf{D}  \\ \hline

\ ASYMFOR-NOF& asymmetric& state of rest  & NL, LI& 400 m, 250 m\\ \hline

\ SYMFOR-NOF& symmetric& state of rest & NL& 400 m, 250 m\\ \hline

\ SYMFOR-SYMJET& symmetric& symmetric jet& NL& 400 m, 250 m\\ \hline

\ SYMFOR-ASYJET& symmetric& asymmetric  jet& NL, LI& 400 m, 250 m\\ \hline

\ ASYMFOR-ASYJET& asymmetric& asymmetric  jet& \begin{tabular}{@{}c@{}}NL \\ LZ, LI \end{tabular} & 400 m\\ \hline

\ \begin{tabular}{@{}c@{}}SYMFOR-MAM \\ SYMFOR-JJA  \end{tabular}& symmetric&\begin{tabular}{@{}c@{}} ERA5 mean flows of MAM or \\ JJA  \end{tabular} & NL& 400 m\\ \hline

\ \begin{tabular}{@{}c@{}}ASYMFOR-MAM \\ ASYMFOR-JJA  \end{tabular}& asymmetric&\begin{tabular}{@{}c@{}} ERA5 mean flows of MAM or \\ JJA  \end{tabular} & NL& 400 m\\ \hline

\end{tabular}
\end{center}
\end{adjustwidth}
\end{table}

\subsection{Energy diagnostics in modal space}

\begin{figure}[ht] 
\includegraphics[width=0.65\textwidth]{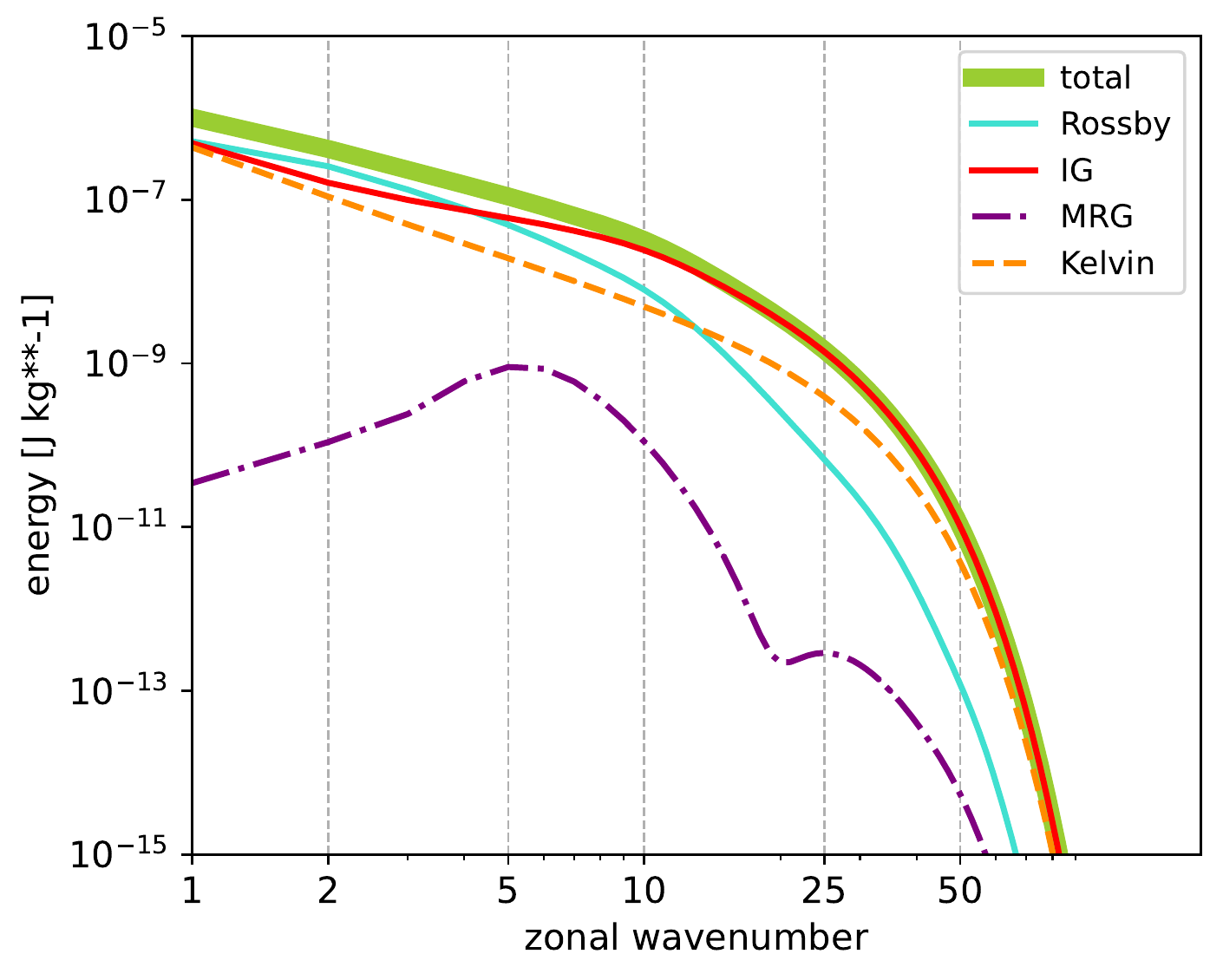}
\caption{Total wave energy spectrum and its partition into contributions from the IG modes, the Rossby modes, the Kelvin mode and the MRG mode. The spectra are a result of SYMFOR-JJA simulation in Table~\ref{t1} run with the JJA background wind of year 1993. The spectra are averaged over 5 days of simulation.}
\label{fig:ener_sp_examp} 
\end{figure}

The energy in $J/kg$ in mode $(k,n,l)$ for $k > 0$ is given by
\begin{equation}
    E_{n,l}^k = gD \left| {\chi_{n,l}^k} \right|^2 \,.
    \label{eq:energy_prod}
\end{equation}
The sum of energy components $E_{n,l}^k$ across all $k$, $n$ and $l$ yields the sum of the kinetic and available potential energy of the system \citep{zagar2015nmd}. Previous studies of the zonal wavenumber energy spectra from reanalyses typically included the MRGW energy among the Rossby wave energy. The comparison of energy levels by MRGWs and other modes by \citet{stephan_zagar2021} showed that the MRGWs contain about 20\% of the westward-propagating non-Rossby wave energy in the troposphere with peak at $k=5-7$. Furthermore, the subseasonal variability associated with the MRGWs makes about $1/3$ of the subseasonal variability in non-Rossby modes. While these numbers are significant, their percentages in the total wave energy (all modes) dominated by the Rossby waves are small.  
As an example, the energy spectra from the SYMFOR-JJA simulation shown in Figure~\ref{fig:ener_sp_examp} demonstrates a nearly realistic energy partition.  Notably, the Kelvin wave energy has a red spectrum and largest values at the smallest wavenumber, whereas the MRG mode energy peaks at synoptic scales in agreement with reanalysis data  \citep{stephan_zagar2021}.

The second diagnostic is targeted at identifying the role of non-linear interactions is the evaluation of the contribution of non-linearities and the forcing to the total energy tendencies. From Equation \ref{eq:energy_prod}, it follows that for $k \geq 1$: 
\begin{equation}
     \frac{d}{dt}E_{n,l}^k \equiv 2gD \ \Re \left[ \left( \frac{d }{d\Tilde{t}} \ \chi_{n,l}^k \right) \left(\chi_{n,l}^k\right)^* \right]= 2gD \ \Re \Bigg\{ \left(f_{n,l}^k - q_{n,l}^k\right) \left(\chi_{n,l}^k\right)^* + \left(q_{n,l}^k \right) \left(\chi_{n,l}^k\right)^* \Bigg\} \,.
    \label{eq:spect_tend}
\end{equation}

Here, the first and the second term on the right-hand side are the energy tendencies by non-linear interactions and the forcing respectively, and $q_{n,l}^k$ denotes the projection of the forcing term on the Hough harmonics. The non-linear tendencies can be further split into interactions between arbitrary wave components, e.g., IG wave - Rossby wave interactions. We evaluate wave-wave interactions between Hough modes for $k > 0$ and wave-mean flow interactions between Hough modes $k>0$ and $k=0$ by applying filters in the non-linear energy tendency in modal space. This diagnostics is particularly useful for studying wave excitations as it allows to separate contributions of various factors to the spectral energy tendency of every mode.

\subsection{Excitation of MRGWs by external forcing}
 
Placing an asymmetric heat source in the tropics leads to a direct excitation of the MRGWs, as the forcing associated with heating directly projects onto the MRG mode, leading to the change in its energy. This experiment is denoted ASYMFOR-NOF in Table~\ref{t1} and it is performed both with linear and non-linear TIGAR using the state of no motion as initial condition. The asymmetric heat source is centred at 2.5°N. 

\begin{figure}[ht] 
\includegraphics[width=1\textwidth]{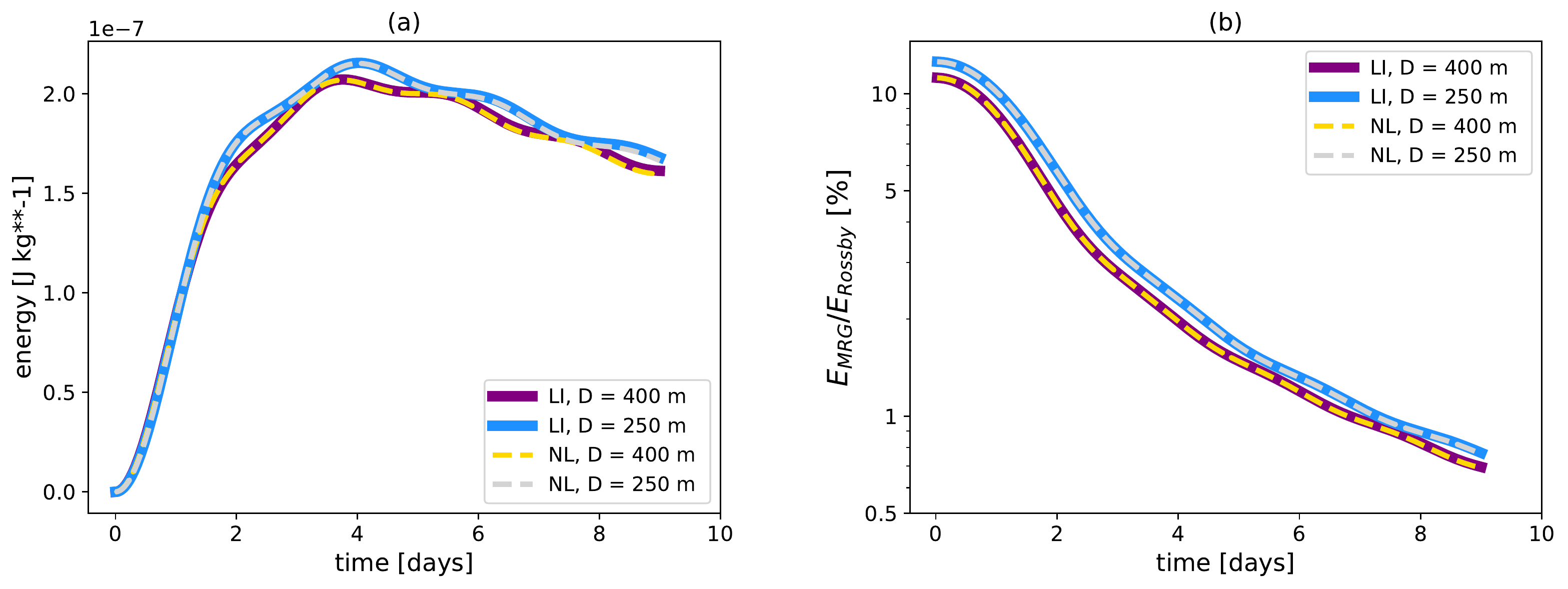}
\caption{(a) Evolution of the MRG mode energy and (b) the evolution of the ratio between the MRG mode energy and the Rossby mode energy in the ASYMFOR-NOF simulations (Table~\ref{t1}). Solid and dashed lines correspond to the simulations in linear (LI) and non-linear (NL) models, respectively. Purple and gold curves denote solutions from the experiments with the mean depth $D=400$ m, whereas blue and silver curves denote the mean depth $D=250$ m.}
\label{fig:lin_exc} 
\end{figure}

Figure~\ref{fig:lin_exc} shows the evolution of the MRG mode energy (panel (a)) and the energy ratio between MRG mode and the Rossby mode (panel (b)) in the linear and non-linear models. It reveals that the MRGWs are excited instantaneously in both models. The energy ratio between the MRG and the Rossby modes is largest initially and decreases gradually until reaching a quasi-equilibrium state depending on the damping timescale (at approximately 4-5 days of simulation time). It is important to notice  that there is no qualitative difference between the linear and non-linear simulations. Thus, wave-wave interactions, which is a proposed MRGW excitation mechanism by \citep{raupp2005}, are irrelevant in this setup.
While a linear excitation of the MRG wave by an asymmetric heat source in the tropics has been known for half a century, its purpose here is to demonstrate how the model and the diagnostics work.

\section{Non-linear excitation of MRGWs in the idealized simulations}\label{sec4}

Now we carry out a number of simulations that provide various details of the MRGW excitation and evolution in the non-linear framework. First, we provide evidence that wave-mean flow interactions can excite MRGWs. Then we concentrate on the scale selection during the excitation process.

Focusing on wave-mean flow interactions, we  exclude the direct excitation of MRGWs by placing the forcing at the equator. The corresponding simulations in Table~\ref{t1} are SYMFOR-NOF, SYMFOR-SYMJET and SYMFOR-ASYJET. 
\begin{figure}[ht] 
\includegraphics[width=0.95\textwidth]{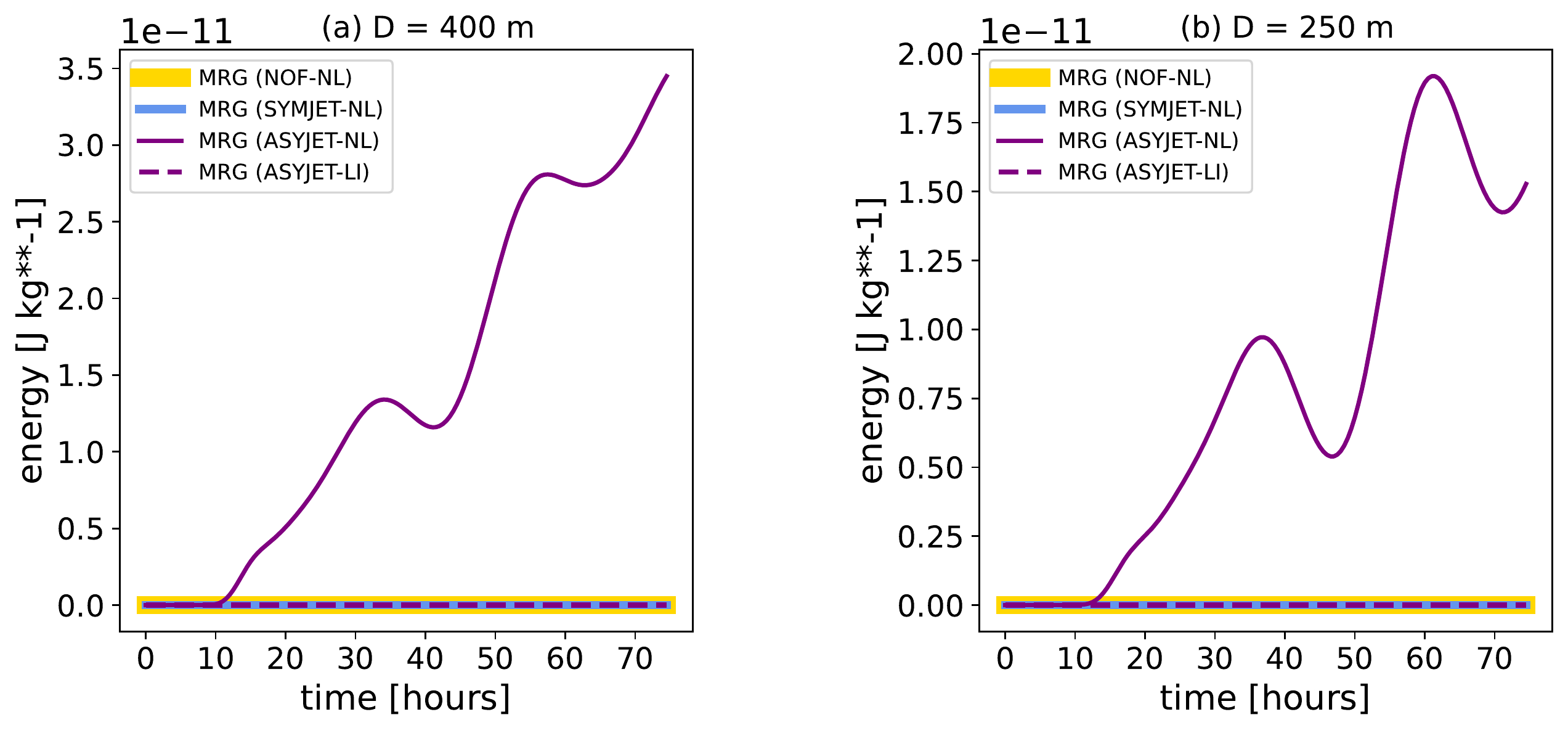}
\caption{Evolution of the MRGW energy in simulations SYMFOR-NOF, SYMFOR-SYMJET and SYMFOR-ASYJET (see Table~\ref{t1} for details) during the first 75 hours of simulation time for mean depths (a) $D=400$ m and (b) $D=250$ m.}
\label{fig:nlin_exc} 
\end{figure}

Figure~\ref{fig:nlin_exc} shows the time evolution of the MRGW energy in the first 75 hours of these three simulations. MRGWs grow only in the case of the non-linear simulation that used an asymmetric background state. This identifies the interactions between the waves generated by the heat source and the asymmetric subtropical jet, i.e., wave-mean flow interactions, as the cause of the MRGW excitation. The fact that at the start of the ASYJET-NL simulation only symmetric $k>0$ modes are present in the system, which cannot produce asymmetric modes by triad interactions, rules out the hypothesis that wave-wave interactions could play a primary role here. Furthermore, the excitation timescale matches the time when the IG waves excited by the prescribed heat source reach the jet region. The excitation timescale in the ASYJET-NL simulations depends on the mean depth $D$ and it is 9 hours and 9.5 hours of simulation time for $D = 400$ m and $D = 250$ m, respectively.

\begin{figure}
    \includegraphics[width=0.99\textwidth]{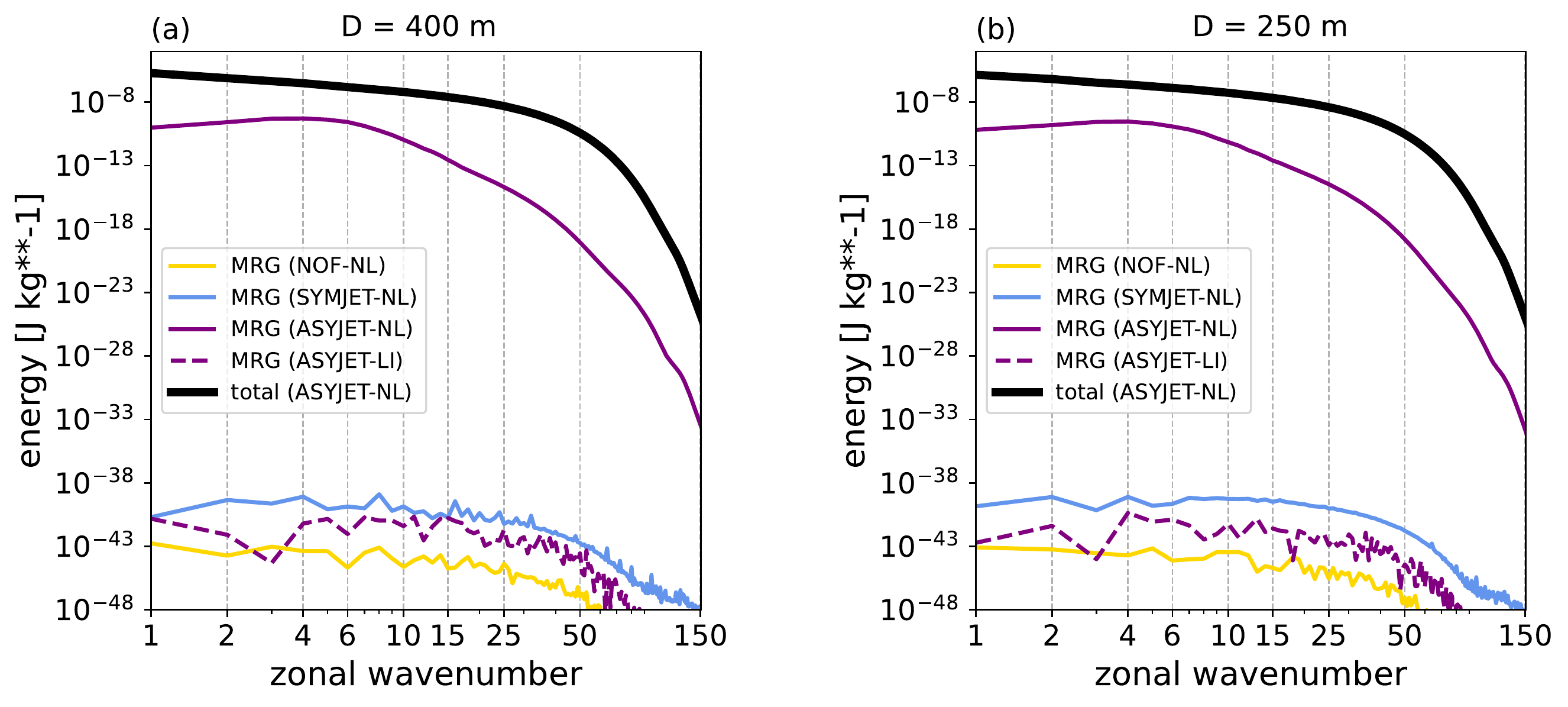}
    \caption{MRGW energy spectra of the SYMFOR-NOF, SYMFOR-SYMJET and SYMFOR-ASYJET (details of simulations in Table \ref{t1}) for (a) $D=400$ m, and (b) $D=250$ m. Only the non-linear SYMFOR-ASYJET simulation produced significant MRGW response. Total wave energy is also shown in the SYMFOR-ASYJET simulations.}
    \label{fig:ener_sp_nlin}
\end{figure}

Figure~\ref{fig:ener_sp_nlin} shows the mean energy spectra of the MRG mode. The averaging window is chosen to be 3 days, which is within the typical lifetime of the observed MRG waves \citep{zangvil1980}. This time window is also much shorter than the time required for vortices to develop within the jet region due to instability. For this reason, instability does not contribute to the MRGW growth within the analyzed window. Figure \ref{fig:ener_sp_nlin} demonstrates the dominance of planetary and synoptic scales in the MRGW response, with a peak in zonal wavenumber  $k = 4$ in the ASYJET-NL simulations in case of both mean depths. The total energy spectrum, with amplitudes 2-3 order of magnitude greater than the MRGW spectrum, contains no peak at large scales.

\begin{figure}
    \includegraphics[width=0.99\textwidth]{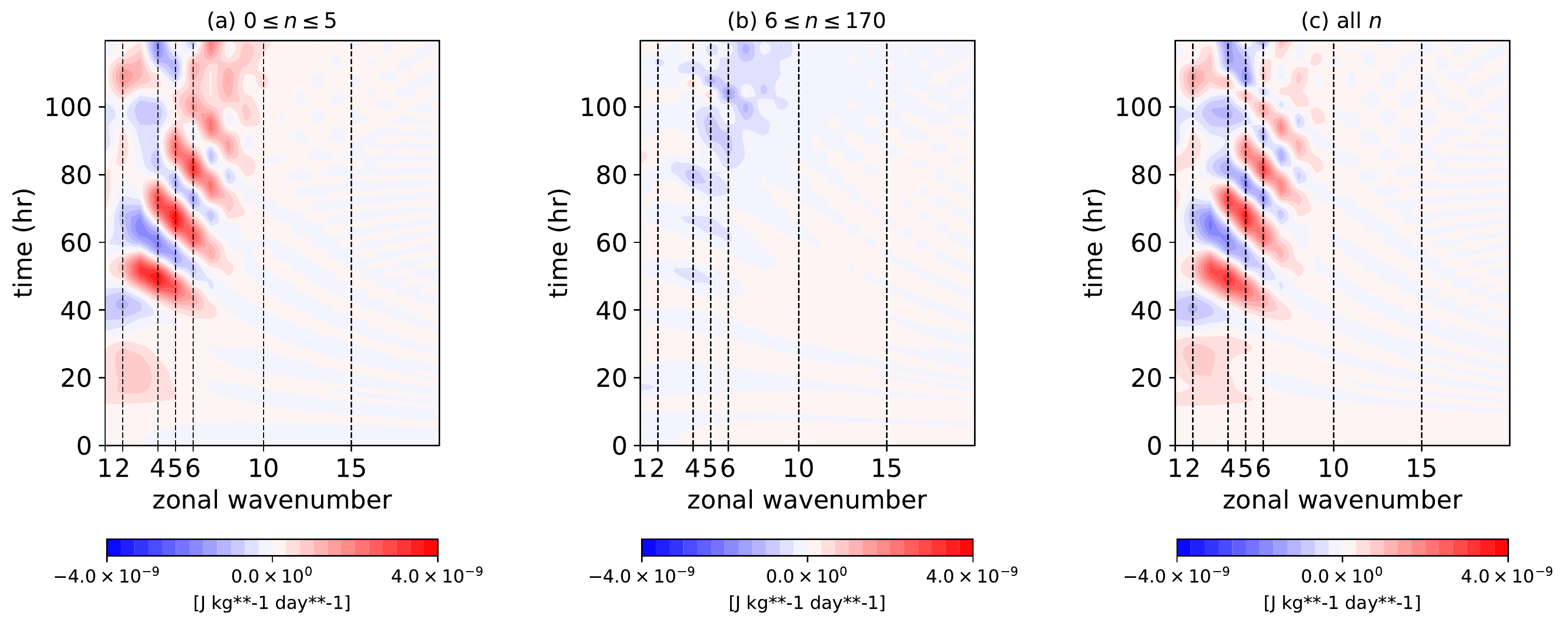}
    \caption{Spectral energy tendencies of the MRG mode as a function of zonal wavenumber and simulation time in the non-linear SYMFOR-ASYJET experiment from Table \ref{t1}. The MRGW energy tendency is partitioned into the the contributions by (a) low meridional modes - mean flow interactions and (b) high meridional modes - mean flow interactions, (c) tendencies due to all non-linear interactions including wave-mean flow and wave-wave interactions.}
    \label{fig:nmodes}
\end{figure}

To generate MRG waves by wave-mean flow interactions, the westerly jet ought to be in relative proximity to the tropics. This is verified by carrying out simulations in which the jets are moved to midlatitudes (i.e., centered at 55° and 60°, not shown). 
In such case, the MRGW energy drops by 5 or more orders of magnitude, and the MRGW energy spectra are white. This is simply explained by the meridional scale of the MRGW and the need for the wave-mean flow interactions to take place within the meridional boundaries of MRGWs. This also means that a crucial role in the MRGW excitation process is played by interaction of the balanced mean state with waves with low meridional wave indices that have their strongest signal on the equatorward side of the jet. This is illustrated in Figure~\ref{fig:nmodes} for the non-linear SYMFOR-ASYJET  simulation with $D$ = 400 m by showing separately the MRGW spectral energy tendencies due to interactions involving low meridional modes $0 \leq n \leq 5$ and modes with high meridional indices $5 \leq n \leq 170$. Low meridional modes are comprised of waves with zonal wind maximum between 30°N and 30°S.
Further separation of tendencies for small $n$ shows that the $n = 1$ Rossby mode provides the largest contribution to the MRGW tendencies although this mode alone can not explain the energy evolution of MRGWs. While IG modes initiate the interactions by reaching the jet the fastest, Rossby mode - mean flow interactions become relevant after about 40 hrs of simulation in case of $D=400$ m, meaning that their contributions to the spectral energy tendencies become comparable to that of IG - mean state interactions.
Similarly, perturbations from midlatitudes or locally imposed perturbations on the jet can excite an MRGW signal depending on the choice of $D$, but the amplitudes of the excited waves are much smaller than in the case of perturbations emanating from tropical heating. 

\begin{figure}[h!]
    \includegraphics[width=0.6\textwidth]{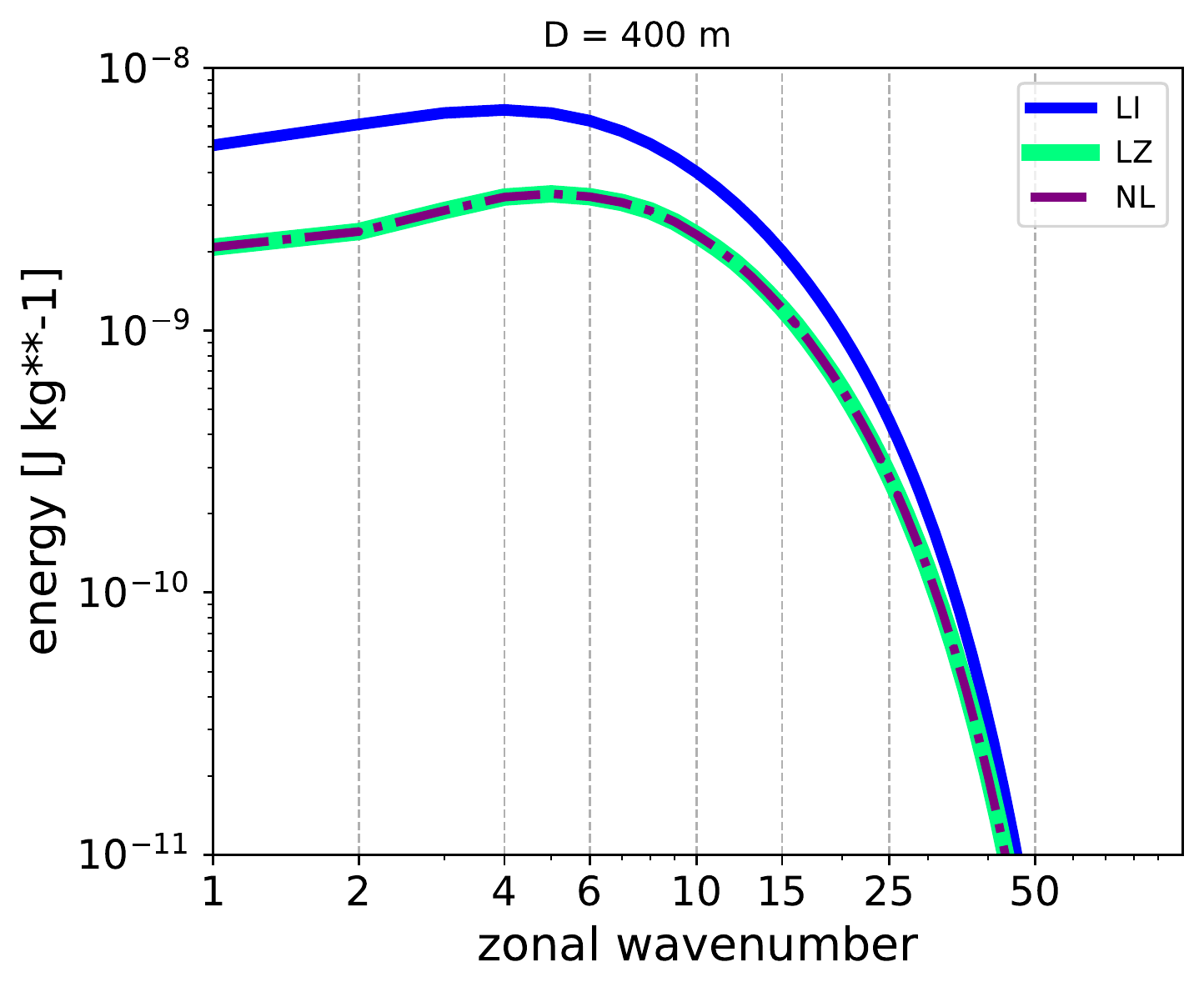}
    \caption{MRGW energy spectra for the linear (LI), linearized (LZ) and non-linear (NL) ASYMFOR-ASYJET simulations. }
    \label{fig:ener_sp_AH_ASYJET}
\end{figure}

The experiments so far showed that the generated MRGWs have the peak signal at large synoptic scales. Here, we further study the scale selection of the waves excited by wave-mean flow interactions. The simulations analysed here include an asymmetric forcing and an asymmetric jet as background flow run with linear, linearized and non-linear TIGAR (i.e., ASYMFOR-ASYJET simulations from Table \ref{t1}). The use of asymmetric forcing ensures that the MRGW is excited in each case, as evident in the MRGW energy spectra shown in Figure~\ref{fig:ener_sp_AH_ASYJET}. 

The MRGW energy spectra in Figure~\ref{fig:ener_sp_AH_ASYJET} are averaged over a 
2-day period. Note that, while the MRGW excitation takes place in all simulations, wave-mean flow interactions act in the linearized and the non-linear model, whereas wave-wave interactions are permitted only in the non-linear simulation. No significant differences can be identified between the MRGW energy spectra of the linearized and the non-linear simulation in Figure \ref{fig:ener_sp_AH_ASYJET} suggesting that wave-wave interactions do not contribute much to the MRGW signal. Furthermore,  all spectra have a spectral peak at $k=4-6$.

\begin{figure}
    \includegraphics[width=0.6\textwidth]{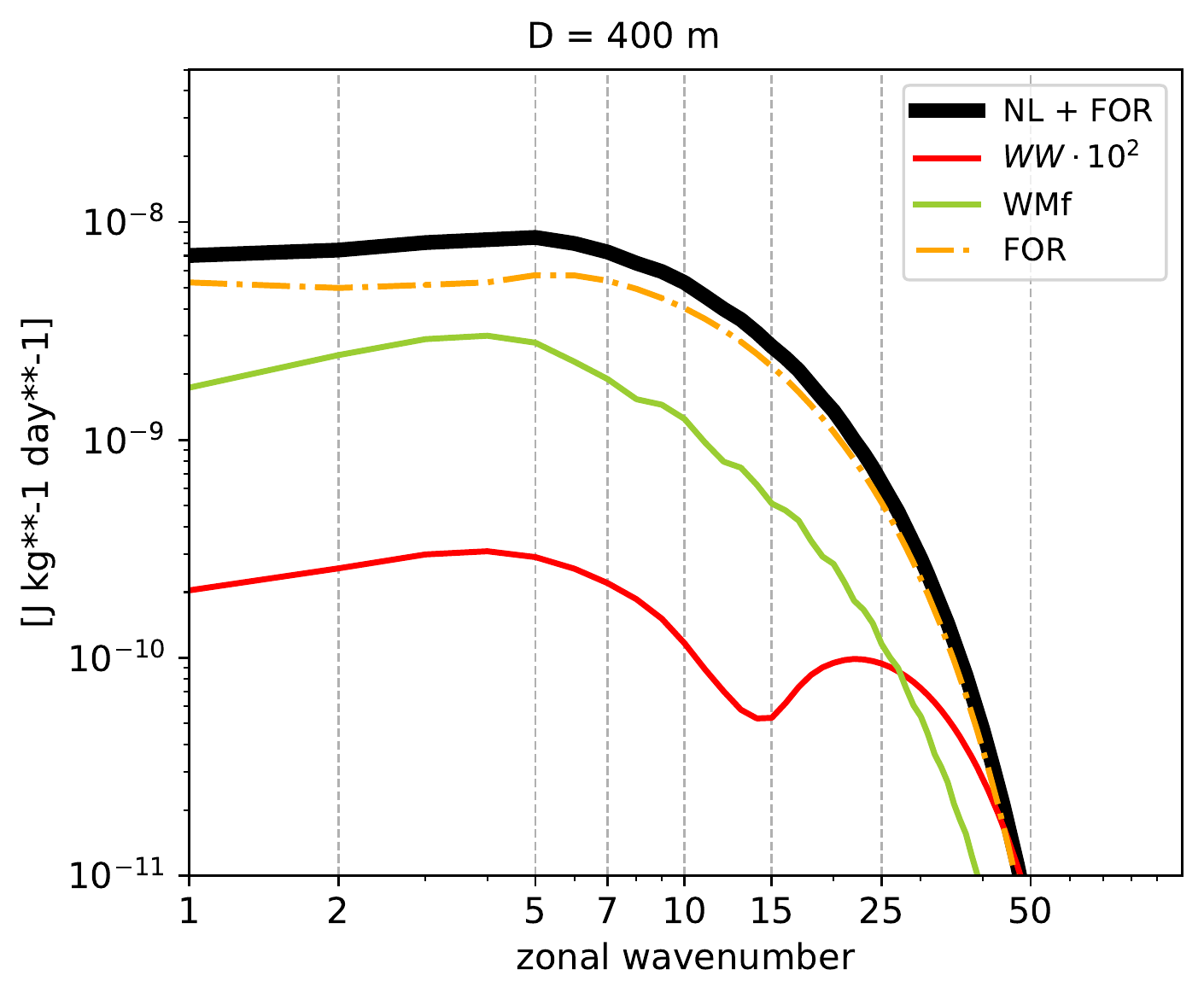}
    \caption{Absolute spectral energy tendencies of the MRG mode  in the non-linear ASYMFOR-ASYJET simulation. The contributions by wave-wave, wave-mean flow interactions and the forcing are denoted WW, WMf and FOR, respectively. NL+FOR stands for the contribution by the non-linear term and the forcing. The WW spectrum is multiplied by $10^2$. Energy tendencies are averaged over two days.}
    \label{fig:sp_tend}
\end{figure}

Focusing now on the role of various processes in the growth of MRGW energy, we analyse the spectra of the absolute energy tendency (Equation~\ref{eq:spect_tend}) in the non-linear simulation. The tendency is partitioned into the contribution by three processes: the external forcing, wave-mean flow and wave-wave interactions. The results are shown in Figure~\ref{fig:sp_tend}. It shows that wave-mean flow interactions play a dominant role in the MRGW scale selection as they are characterized by a well-defined spectral peak that is located at $k=3-5$. 
The forcing, which is Gaussian, makes the total spectrum more flat at scales $k<15$ and its amplitude for the selected model setup exceeds that of the wave-mean flow interactions. The contribution of wave-wave interactions is approximately 3 orders of magnitude smaller than that of the other two mechanisms.  
 
 The observed spectral peak in the MRGW energy spectra is somewhat sensitive to the averaging period and evolves with time (not shown). Nevertheless, we find that local maxima in the MRGW energy wave spectra are always attributed to the wave-mean flow interactions. Furthermore, as the mean depth rises, the relative importance of wave-mean flow interactions increases and can even exceed the amplitude of energy tendencies due to the forcing at large scales.

\section{Excitation of MRGWs in simulations with realistic background flow}\label{sec5}

Now we apply background zonal-mean zonal winds derived from ERA5 for the boreal spring and summer seasons, as listed in Table \ref{t1}. Our goal is to verify to what extent the conclusion based on the idealized flows carry over to a more realistic setting and to quantify the role of observed asymmetry on the amplitude of the MRGWs. As discussed in section 2, the JJA season has a higher asymmetry than the MAM season.

\begin{figure}
    \includegraphics[width=0.99\textwidth]{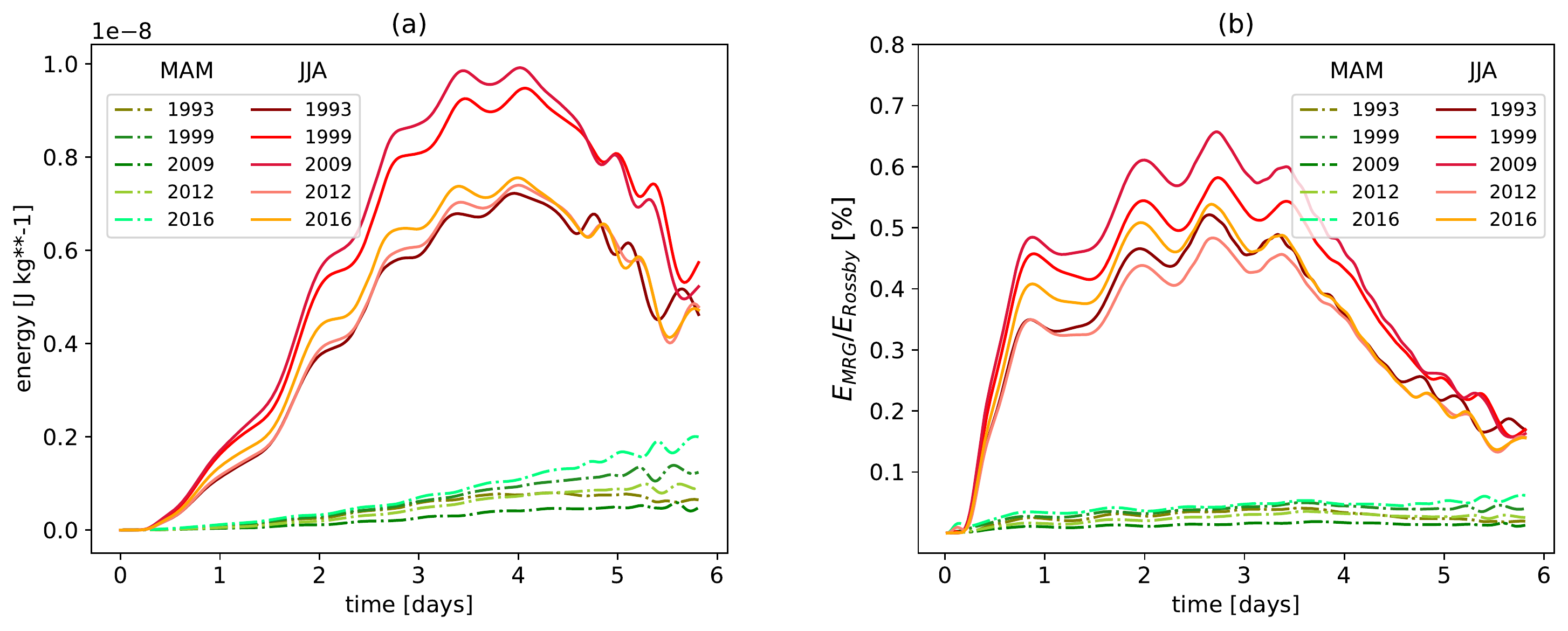}
    \caption{ a) MRGW energy evolution and b) evolution of the ratio between the energy of the MRG mode and the Rossby mode (expressed in \%) in the tropical heat source simulations using realistic background flows. The shades of red and green correspond to the simulations with the JJA and MAM background flows, respectively.}
    \label{fig:real_flow_MRG}
\end{figure}

First, we consider simulations in which the excitation by the external forcing is suppressed, i.e., the forcing is symmetric about the equator. The plots in Figure~\ref{fig:real_flow_MRG} show the evolution of the MRG mode energy for the two seasons and the evolution of the ratio between the MRG and the Rossby mode energy.

\begin{figure}
    \includegraphics[width=0.55\textwidth]{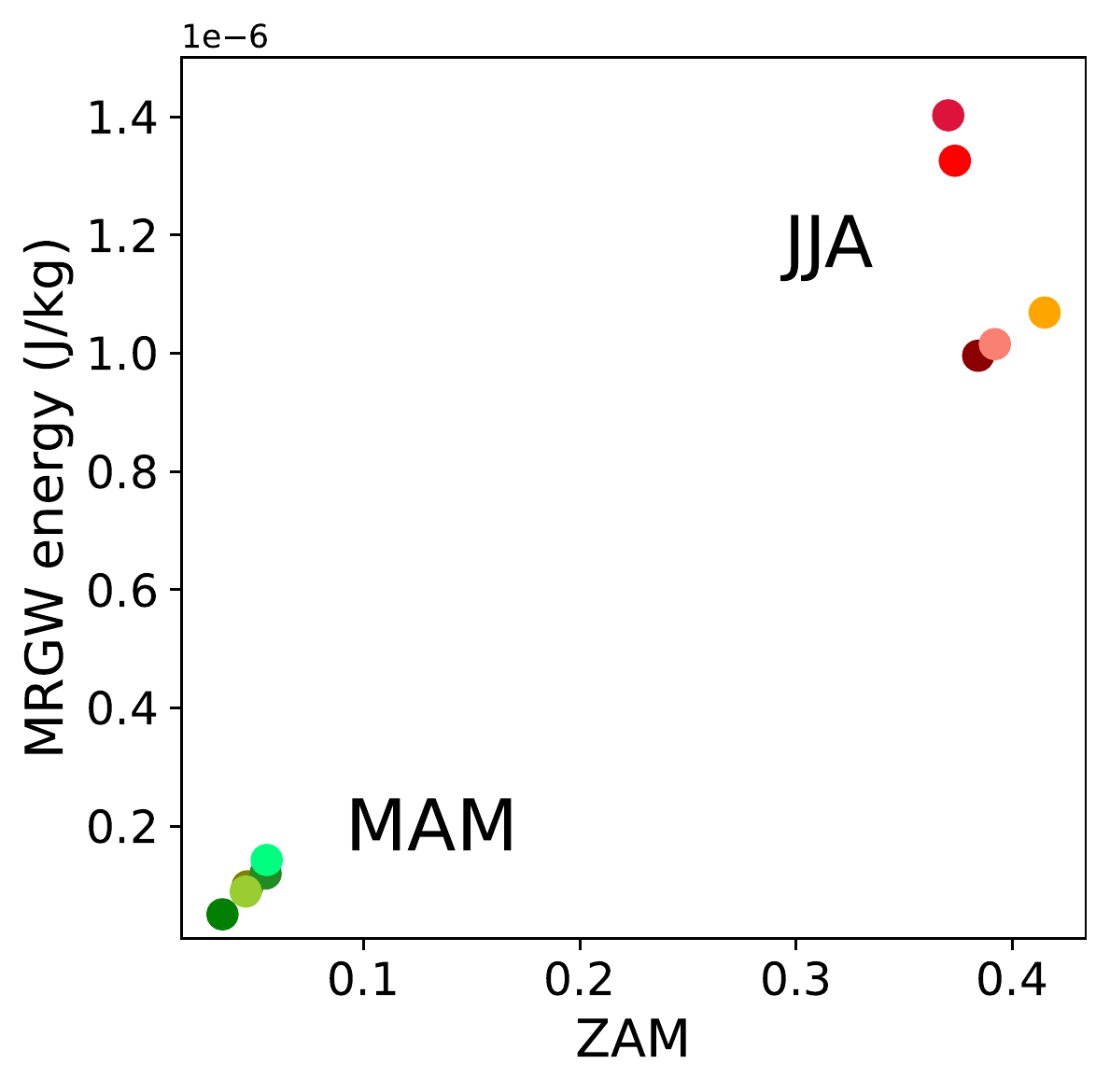}
    \setlength\abovecaptionskip{0.7\baselineskip}
    \caption{MRGW energy integrated over 5 days of simulation as a function of the asymmetry measure (ZAM) defined in Section \ref{setup} The shades of red and green represent JJA and MAM simulations respectively.}
    \label{fig:scatter}
\end{figure}

Figure~\ref{fig:real_flow_MRG} reveals large differences in the MRGW energy evolution and its relative percentage between the simulations with the JJA and MAM seasons, with substantially larger values for JJA. For example, the maximum of the energy of the MRG mode on day 4 in the JJA simulations is approximately 5 times larger compared to the simulations with MAM background flows. The maximum ratio between the energy of the MRG mode and the Rossby mode is approximately 0.65\% in the JJA simulations, whereas it stays below 0.08\% in the MAM simulations. We note that we did not find any major difference between the stability profiles of the JJA and MAM flows and we did not observe barotropic instability development during these simulations, therefore the MRGW excitation is only a result of wave-mean flow interactions. 
The 5-day averaged MRGW energy for every experiment is presented as a function of ZAM, the zonal asymmetry measure, in Figure~\ref{fig:scatter}, which provides a clear evidence that the larger the zonal asymmetry, the more energetic the MRG waves are. The JJA and MAM simulations appear as two clusters in Figure~\ref{fig:scatter}.

\begin{figure}[h!]
    \includegraphics[width=0.99\textwidth]{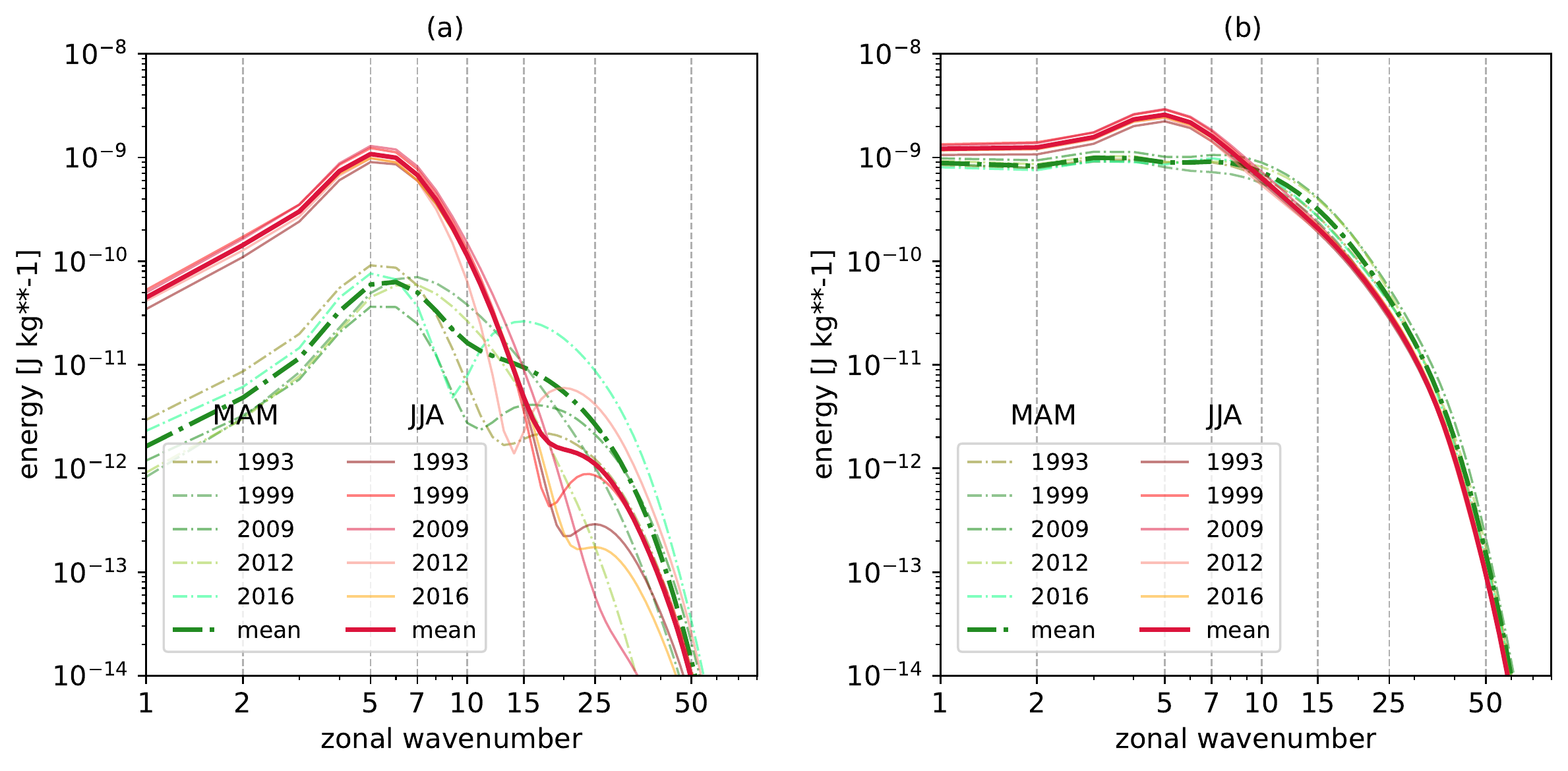}
    \setlength\abovecaptionskip{0.85\baselineskip}
    \caption{Energy spectra of the MRGWs in simulations with realistic background flows run with a) symmetric and b) asymmetric forcing. The averaging period is 5 days. Thin and thick curves denote the individual simulations and the means over JJA and MAM simulations, respectively.}
    \label{fig:en_spectra_era5flows}
\end{figure}

The mean energy spectra of the MRGWs in the simulations with realistic background flows is shown in Figure~\ref{fig:en_spectra_era5flows} for the case of both symmetric and asymmetric forcing. The averaging period is 5 days. We find that even with the excitation by the forcing, simulations with more asymmetric background wind (JJA) produce more energetic MRG waves at large and at synoptic scales (i.e. up to $k$ = 9 and 14 in case of the asymmetric and the symmetric forcing respectively). At the scales $k = 3 - 7$, the JJA simulations yield 10 and 3 times larger MRGW energy in case of symmetric and asymmetric forcing, respectively.  The differences between the JJA and MAM spectra vanish as the zonal wavenumber increases towards $k=50$.

The forcing symmetry plays a role for the MRGW scale selection. In case of a symmetric forcing, the dominant MRGW wavenumber is in range of $k=5-7$ for both MAM and JJA background flows. When an asymmetric forcing is applied, the mean MAM energy spectrum flattens without showing any prominent peaks, whereas the mean JJA spectrum has a well-defined peak in range of $k$=3-7. This result implies that the spectral tendencies of the MRGW energy due to wave-mean flow interactions, which are responsible for the MRGW scale selection, are smaller than tendencies due to forcing when the mean flow is less asymmetric.

\section{Discussion and Conclusions}\label{sec6}

The underlying mechanism responsible for the excitation of mixed Rossby-gravity waves (MRGWs) in the tropical atmosphere is poorly understood, particularly concerning their synoptic scale selection, which cannot be fully explained by the excitation associated with the tropical heating \citep{holton1972waves_heatsrc,kosaka2005}. To address this significant knowledge gap, we employ a new forecasting model TIGAR. This model incorporates the MRG mode as a prognostic variable alongside other waves. Such an approach enables a direct quantification of various factors contributing to MRGW development within the model, presenting a distinct advantage over conventional equatorial wave filtering techniques.

Using this novel framework, we first demonstrate that MRGWs can be exited by an asymmetric tropical heat source. 
Then we show that MRGWs are excited by wave-mean flow interactions provided that the mean flow is asymmetric with respect to the equator.
In the performed simulations, waves initiating the interactions are generated by a heat source at the equator. The excited MRGWs have the maximal energy at the zonal wavenumbers $k=4-5$, in agreement with observations \citep[e.g.][]{yang2017,stephan_zagar2021}. The non-linear wave-mean flow interactions as an excitation and scale selection mechanism is further confirmed through the analysis of the spectral energy tendencies. Depending on the prescribed mean fluid depth $D$ and the time stage of the process, the wave-mean flow interactions are mainly due to waves with low meridional wave index (e.g. for $ D = 400$ m, the $n=1$ Rossby mode is the main contributor to the MRGW growth).

\begin{figure}[b!]
    \includegraphics[width=0.65\textwidth]{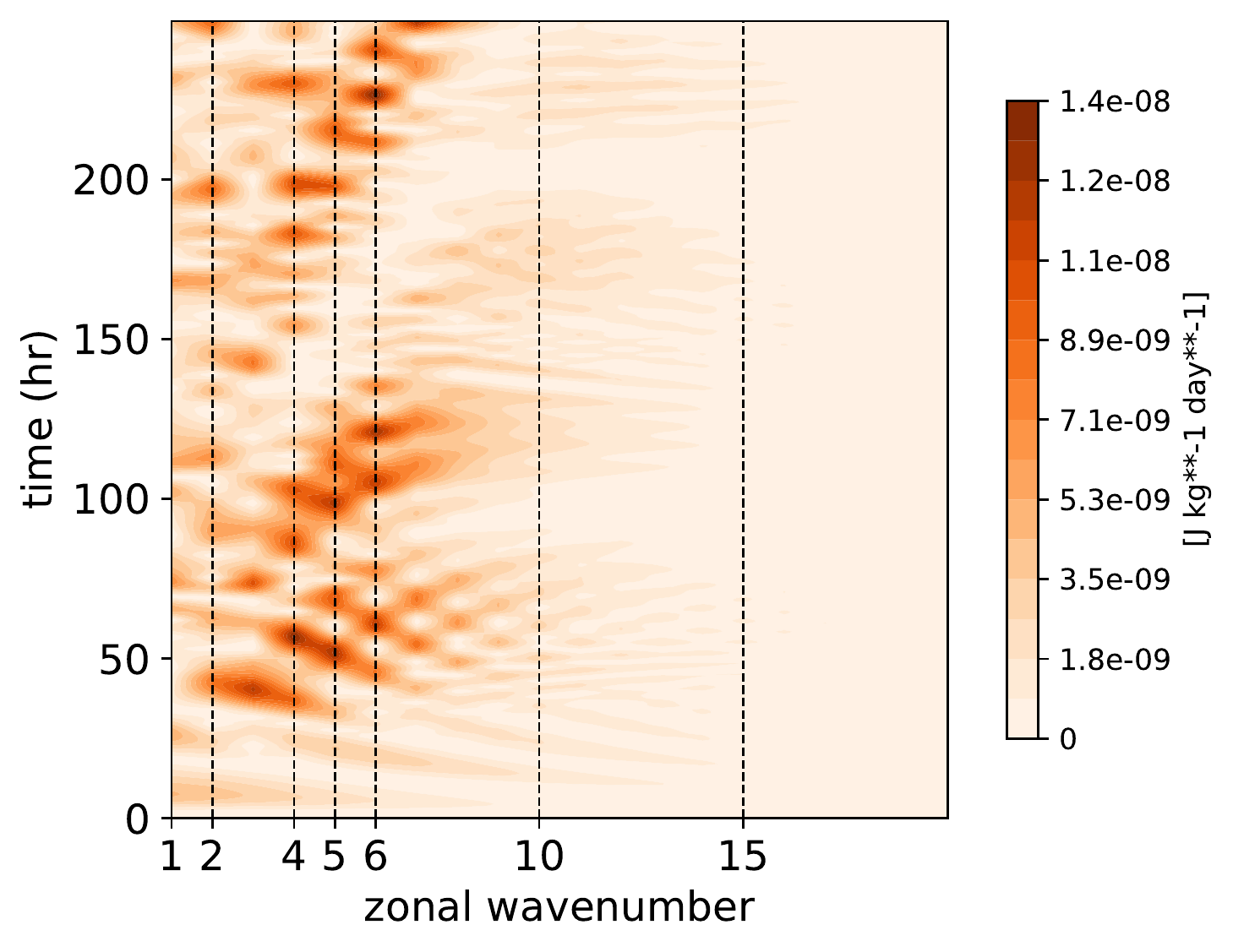}
    \setlength\abovecaptionskip{0.9\baselineskip}
    \caption{Absolute spectral energy tendencies of the MRGW as a function of zonal wavenumber and time in the non-linear ASYMFOR-JJA simulation with the zonal mean zonal wind of JJA 1993.}
    \label{fig:MRG_en_tend_zw_t}
\end{figure}

The background flow itself can potentially determine the selected zonal wavenumber of the excited MRG waves through the instability growth associated with linearized dynamics. This hypothesis is verified by studying the eigenmodes of the system linearized about the pertinent background state. The background eigenmodes are determined following the algorithm suggested by  \citet{kasahara1980}. We find unstable background eigenmodes with substantial projection onto the MRG modes in range of $k=1-17$ and $k=1-19$ for the idealized and realistic background zonal flows respectively. However, most of these unstable eigenmodes have a small growth exponent, which varies only slightly across nearby scales. For this reason, the selected MRGW scales generally do not coincide with the fastest growing mode of the linearized system, rather they are determined by the peak in the wave-mean flow interactions. This peak fluctuates in time due to the change in phases of interacting waves, which explains the slight movement of the MRGW spectral peak. For instance, Figure~\ref{fig:MRG_en_tend_zw_t} shows the amplitudes of the contributions of the wave-mean flow interactions to the MRGW tendencies over 10 days of a simulation with a realistic background flow in JJA 1993. It shows that the preferred zonal wavenumber, which is a recipient of most of energy from interactions, changes over time, but remains at zonal wavenumbers between $k=3$ and $k=6$, which is consistent with the maximal MRGW signal  at $k=5$ in the reported simulations. 

The contribution of the forcing to MRGW tendencies is found to be nearly flat for the smallest wavenumbers, whereas the MRGW tendencies by wave-mean flow interactions, which involve waves produced by the forcing, fluctuate over a small range of $k$ as the phases of these waves change in time. Should the forcing have a peak zonal wavenumber in the MRGW absolute tendency spectra, wave-mean flow interactions would also peak at the same $k$, therefore the scale of the generated MRGWs would be completely predictable. The peak scale of the MRGW excitation is likely sensitive to waves in the background, which is a potential subject of a future study.

Two earlier studies of non-linear excitation mechanism of MRGWs require discussion. First, \citet{itoh1988} showed that asymmetric lateral boundary forcing of the thermodynamic equation reinforces asymmetric tropical modes. They suggested that there may be enhanced MRGW activity in the summer and winter season when the midlatitude westerlies in the winter hemisphere are  
intensified. They proposed wave-CISK MRGW excitation mechanism, which introduces asymmetry through boundary forcing and, therefore, requires certain properties of extra-tropical flows to maintain. Here we demonstrate a different mechanism, where MRGW growth is promoted by wave-mean flow interactions that are local to the tropics and do not require any extratropical input. 

Second, we could not find a strong evidence of the excitation of MRGWs by wave-wave interactions suggested by Raupp and Silva Dias (2005). In their simulations with an equatorial $\beta$-plane shallow water model, the initial state was prepared using an off-equatorial stationary heat source and they observed strong MRGW development, provided IG waves were suppressed in the model. In our experiments, an asymmetric heat source leads to the growth of MRGWs directly regardless of the presence of the IG waves. Moreover, suppressing IG modes in our simulations does not significantly alter the contribution of wave-wave interactions to MRGW tendencies (not shown); i.e. wave-wave interactions are negligible compared to the wave-mean flow interactions and the excitation process by the external forcing.

Although our simulations are performed in a simplified model, quantitatively they are consistent with the MRGW energy estimates in reanalysis data. For instance, an intercomparison of the MRGW energy levels in four re-analyses datasets by \citet{zagar2009uncertainties_partii} showed that they comprise between 4\% and 15\% of the non-Rossby wave flow  (IG+MRG+Kelvin wave).
These percentages roughly correspond to values obtained in simulations with a realistic background flow in the present study. In contrast to its energy, the share of the MRGW signal in intraseasonal  variance is large \citet[][]{stephan_zagar2021}.

The atmosphere is never in a symmetric zonal mean state. We demonstrated that the level of asymmetry, defined by a new spectral asymmetry measure (Equation \ref{eq:ZAM}), determines the amplitude of the MRGW signal. 
In particular, the simulations with the zonal mean state from the JJA season in ERA5 reanalyses produce a significantly stronger MRGW response in large and synoptic scales than the ones with less asymmetric zonal wind profiles from the MAM season. For example, the MRGWs in JJA simulations are roughly 10 times more energetic compared to MAM simulations when a symmetric forcing is applied.

Similarly to the background flow, the real atmosphere is asymmetric also in the distribution of convective forcing with the greatest asymmetry during boreal summer. The quantification of the relative importance of the excitation by the external forcing (that mimics convection) and by wave-mean flow interactions in the real atmosphere requires the extension of the present study to the three-dimensional moist atmosphere.

We speculate that the differences between tropospheric and stratospheric MRGWs such as the zonal scale with the largest signal could be explained by wave-mean flow interactions that may peak at different scales compared to the simulations run with tropospheric flow profiles. The MRGW filtering in the operational ECMWF forecasts at the MODES webpage, http://modes.cen.uni-hamburg.de shows much larger scales and greater amplitudes of the MRGWs in the middle atmosphere compared to the troposphere. This may be a result of the larger asymmetry of the zonal mean wind, i.e. an enhanced MRGW generation by wave-mean flow interactions in the middle atmosphere, however, this is a subject of a follow-on study.

Finally, we note that the presented results can help to better understand the role of MRGWs in tropical intrinsic and practical predictability and tropical-extratropical interactions. The non-linear excitation process suggests a limited intrinsic predictability of tropical synoptic-scale disturbances with a significant meridional circulation component.


\section*{Acknowledgments}
This paper is a contribution to the project W6 (Spectral Energy Fluxes by Wave-Wave interactions) of the Collaborative Research Centre TRR 181 "Energy Transfers in Atmosphere and Ocean" funded by the Deutsche Forschungsgemeinschaft (DFG, German Research Foundation) - Projektnummer 274762653. We would like to thank Rolando Garcia for the discussion and the useful comments that aided this manuscript. We further acknowledge the discussions with Frank Lunkeit, who contributed to the improvement of the paper.

\nolinenumbers

\bibliography{library}

\bibliographystyle{agsm}

\end{document}